\documentclass[jap,twocolumn, showpacs,preprintnumbers,amsmath,amssymb,superscriptaddress,floatfix]{revtex4} 
\usepackage[]{graphicx}
\usepackage{graphics}
\usepackage{subfigure}
\usepackage{amsmath}
\usepackage{tabularx}
\usepackage{bm}
\usepackage{multirow}
\begin{document}

\title[Tuning the Fano resonance between localized and propagating surface plasmon resonances for refractive index sensing applications]{Tuning the Fano resonance between localized and propagating surface plasmon resonances for refractive index sensing applications}

\author{Kristof Lodewijks}
\email[Corresponding author: ] {lode@kristoflodewijks.be}
\affiliation{IMEC, Kapeldreef 75, B-3001 Leuven, Belgium}
\affiliation{Department of electrical engineering (ESAT), KU Leuven, Leuven, Belgium}

\author{Jef Ryken}
\affiliation{IMEC, Kapeldreef 75, B-3001 Leuven, Belgium}
\affiliation{Department of Biosystems (MeBioS), KU Leuven, Leuven, Belgium}

\author{Willem Van Roy}
\affiliation{IMEC, Kapeldreef 75, B-3001 Leuven, Belgium}

\author{Gustaaf Borghs}
\affiliation{IMEC, Kapeldreef 75, B-3001 Leuven, Belgium}
\affiliation{Department of Physics and Astronomy, KU Leuven, Leuven, Belgium}

\author{Liesbet Lagae}
\affiliation{IMEC, Kapeldreef 75, B-3001 Leuven, Belgium}
\affiliation{Department of Physics and Astronomy, KU Leuven, Leuven, Belgium}

\author{Pol Van Dorpe}
\email[Corresponding author: ] {pol.vandorpe@imec.be}
\affiliation{IMEC, Kapeldreef 75, B-3001 Leuven, Belgium}
\affiliation{Department of electrical engineering (ESAT), KU Leuven, Leuven, Belgium}

\begin{abstract}
Localized and propagating surface plasmon resonances are known to show very pronounced interactions if they are simultaneously excited in the same nanostructure. Here we study the fano interference that occurs between localized (LSPR) and propagating (SPP) modes by means of phase sensitive spectroscopic ellipsometry. The sample structures consist of periodic gratings of gold nanodisks on top of a continuous gold layer and a thin dielectric spacer, in which the structural dimensions were tuned in such a way that the dipolar LSPR mode and the propagating SPP modes are excited in the same spectral region. We observe pronounced anti-crossing and strongly asymmetric line shapes when both modes move to each others vicinity, accompagnied of largely increased phase differences between the respective plasmon resonances. Moreover we show that the anti-crossing can be exploited to increase the refractive index sensitivity of the localized modes dramatically, which result in largely increased values for the Figure-Of-Merit which reaches values between 24 and 58 for the respective plasmon modes.
\end{abstract}

\maketitle

Surface plasmon resonances have been widely studied over the last decades for different biological and chemical sensing applications \cite{NatMaterAnker}. Both propagating Surface Plasmon Polaritons (SPPs) and Localized Surface Plasmon Resonances (LSPRs) exhibit very interesting properties for sensing applications \cite{NLSvedendahl2009,OptExprPiliarik} due to their high degree of tunability and their susceptibility to the dielectric properties of the surrounding environment. Moreover, the dimensions of many structures that support (localized) plasmon resonances are very similar to the scales of (biological) molecules, which makes them an ideal interface medium. The most widely studied applications include Surface Enhanced Raman Scattering (SERS) \cite{APLYe, NanoLettYe} and Refractive Index (RI) sensing \cite{NatMaterAnker,NLSvedendahl2009,ACSOffermans,NanoTechChen}. In the field of refractive index sensing many research group have focussed on plasmon line shape tuning in order to narrow down the line widths, which results in a higher Figure-Of-Merit (FOM $= (d \lambda / d n)/fwhm$), which makes it possible to reach lower detection limits. One path to reach this goal is to look at Fano interference between different plasmonic modes \cite{ACSFrancescato,ACSSvedendahl,NanoLettVerellenDolmen} which has been applied succesfully for refractive index sensing applications \cite{ACSHao,NanoLettVerellenCross,NanoLettHao,NatMaterLulyanchuk,NanoLettEvlyukhin}. It was also shown before that measuring the phase instead of the amplitude of SPPs \cite{OptCommKabashin,OptExprKabashin} and LSPRs \cite{NanoLettLodewijks,OptLettKravets} significantly reduces the line widths of the plasmon modes with up to 2 and 1 order of magnitude respectively. It is also well established that periodic arrays of localized plasmon resonators show strong coupling effects, which also allows to tune their resonance line width and position \cite{ACSOffermans}. If periodic arrays of localized plasmon resonators are positioned in the vicinity of a metal layer that supports propagating SPP modes then these modes tend to show pronounced coupling to the propagating modes and anti-crossing behavior between LSPR and SPP modes is observed \cite{OptLettChu,APLGhoshal,PRBChrist}. 

In this work, we investigate the interaction between LSPR and SPP modes by means of spectroscopic ellipsometry. We observe strong Fano interference between both types of plasmon modes, which is reflected in highly asymmetric line shapes and a pronounced increase of the phase difference between the different plasmon modes. We show how this interference can be exploited to increase the refractive index sensitivity of the LSPR mode by controlling its spectral position through the SPP mode. By measuring the phase instead of the amplitudes of the resonances, we manage to reduce the line widths of the plasmonic modes significantly, resulting in extremely high values for the sensing FOM.The investigated sample structures are illustrated in figure 1 (panel a and b) and consist of a periodic array of gold nanodisks on top of a 50 nm silica spacer layer and a continuous gold layer. The pitch in both directions is fixed at 400 nm, while the diameter of the gold nanodisks is 100 and 140 nm. The periodicity of the sample structures offers two key advantages: (1) The line width of the LSPR modes is reduced as the effects of inhomogeneous broadening are largely suppressed and the coupling between the neighboring particles is identical for all nanostructures; (2) The array of nanoparticles act as a grating structure which allows for very efficient excitation of propagating SPP modes on the gold layer below. The diameter of the gold disks was chosen to obtain overlap between the spectral positions of the SPP and LSPR modes, such that their interactions could be investigated. We performed angle- and polarization dependent spectroscopic ellipsometry measurements \cite{SOPRA} in reflection mode in order to obtain access to both phase and amplitude of the SPP and LSPR modes \cite{NanoLettLodewijks,OptLettKravets}, which allows us to study their interactions in more detail. To do so, the polarization of the incident wave is modulated between P- and S-polarization and the phase information is extracted by performing lock-in measurements at the modulation frequency. The measured quantities $\tan(\Psi)$ and $\cos(\Delta)$ are related by the main equation of ellipsometry: 
\begin{equation}
\begin{split}
\rho = & \frac{R_{P}}{R_{S}} = \tan(\Psi) \exp(i \Delta) \\
  & = \tan(\Psi) (\cos(\Delta) + i \sin(\Delta))
\end{split}
\end{equation} 

and represent the amplitude reflection ratio between P and S $(\tan(\Psi))$ and the phase difference between the reflected signals $\Delta$ for the 2 polarizations (reflected in the $\cos(\Delta)$ value). 

The gold nanodisks were fabricated by conventional e-beam lithography (EBL) on PMMA resist and subsequent evaporation of gold nanodisks. After the sample fabrication a short annealing step was applied in order to reduce the damping of the plasmon modes \cite{NanoLettChen} (more details on sample fabrication in supporting information S1).  The electromagnetic angle- and polarization dependent response of the samples was modelled in the RF module of  COMSOL Multiphysics \cite{COMSOL} for one unit cell (wired box in figure 1 a) using periodic Bloch boundary conditions. In that way both the amplitude and the phase of the reflected waves could be extracted by averaging the complex fields of the reflected waves over one unit cell. 

\begin{figure}
\includegraphics[width=8.7cm]{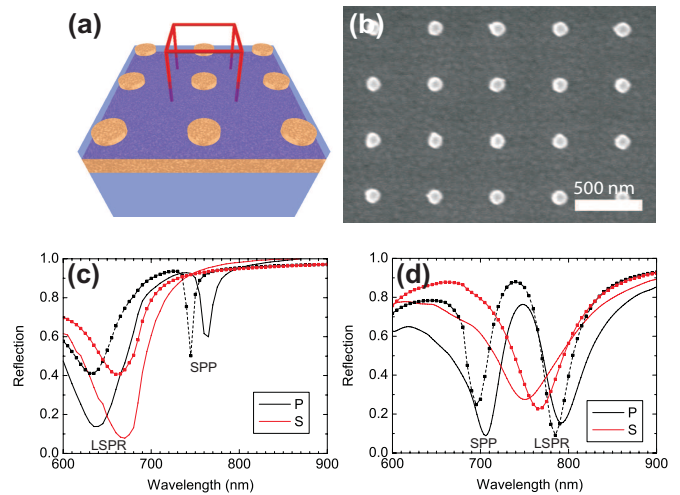}
\caption{(a) Schematic overview of the periodic nanodisk samples. The pitch is 400 nm in both directions and the disk diameter is fixed at 100 nm and 140 nm. The red wire box indicates the unit cell used in simulations. (b) Scanning Electron Microscope picture of one of the nanodisk samples. (c) and (d) Measured (full lines) and simulated (dahsed lines) reflection spectra for 100 nm and 140 nm disks respectively at an incident anlge of $45^o$ in P- and S-polarization. Note that in both cases an SPP grating mode is excited for P-polarized light and the dipole LSPR modes are excited for both polarization states. For 100 nm case the SPP and LSPR mode in P-polarization show little or no interaction, while in the 140 nm case the interaction of the modes results in an asymmetric line shape.}
\label{figure1}
\end{figure}

Panels c and d of figure 1 show the measured and simulated reflection spectra at an angle of incidence of $45^o$ for the 100 and 140 nm disk samples respectively. For both samples we observe a pronounced LSPR mode in both polarization states and a propagating SPP mode for the P-polarization. For 100 nm disks (panel c) in P-polarization the LSPR mode is observed at shorter wavelengths than the SPP mode, while both of them are spectrally well separated. They show little or no interaction, and as a result their line shape is highly symmetric. For the 140 nm disks (panel d) the SPP mode is observed at shorter wavelengths than the LSPR mode and they show more spectral overlap. Therefore, these modes tend to interact more pronounced, resulting in asymmetric Fano line shapes. In a next step we investigated the angle dependence of the different plasmon modes and their interactions. Figure 2 shows an overview of the measured and simulated angle- and polarization dependent reflection spectra for 100 nm disks in air. 

\begin{figure}
\includegraphics[width=8.7cm]{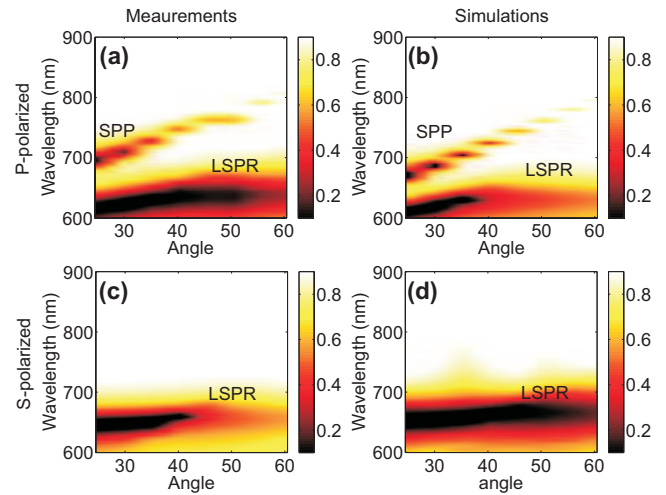}
\caption{Intensity plots for angle dependent reflection measurements and simulations on 100 nm disks in air. (a) Measured spectra for P-polarization. (b) Simulated spectra for P-polarization. (c) Measured spectra for S-polarization. (d) Simulated spectra for S-polarization. In P-polarization we clearly observe the -1 diffracted SPP mode that shows little or no interaction with the dipolar LSPR mode, while in S-polarization we only observe the dipolar LSPR mode.}
\label{figure2}
\end{figure}

For P-polarized waves (panels a and b) the LSPR mode is observed around 620 nm for small angles of incidence and shows a minor red shift up to 650 nm as the incident angle is increased. The SPP mode is observed at longer wavelengths and shows a pronounced red shift from about 700 nm to 800 nm as the incident angle is increased. Therefore at small angles of incidence we observe a minor interaction between the LSPR and SPP mode, resulting in slightly asymmetric line shapes. As the grating SPP mode shifts to longer wavelengths, the interaction between both modes is reduced when the incident angle increases. The observed grating mode is the $\nu = -1$ diffracted order which can be described by the grating formula
\begin{equation}
k_{spp} = k_0 sin \theta \pm \nu \frac{2 \pi}{a}
\end{equation} 
in which $k_0$ is the incident wave vector, $\theta$ the incident angle and $\nu$ the diffracted order. The spectral shift with increasing incident angle can easily be understood from the dispersion relationship of the SPP mode, as illustrated in the supporting information (S2). The $\nu = +1$ diffracted order is not observed in our experimental spectra, as it is expected to be observed at higher energies where the reflection spectra are dominated by absorption due to interband transitions in the gold. For this mode a blue shift with increasing angle of incidence is expected, which could be observed for larger grating pitches. For S-polarized waves we only observe the LSPR mode which shows a minor red shift with increasing incident angles. For 140 nm disk samples we observe a quite different behavior both in measurements and simulations, as depicted in the spectral plots in figure 3.  

\begin{figure}
\includegraphics[width=8.7cm]{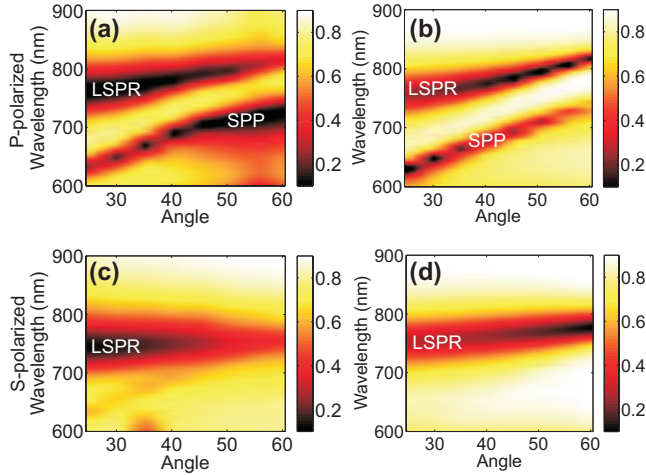}
\caption{Intensity plots for angle dependent reflection measurements and simulations on 140 nm disks in air. (a) Measured spectra for P-polarization. (b) Simulated spectra for P-polarization. (c) Measured spectra for S-polarization. (d) Simulated spectra for S-polarization. In P-polarization we clearly observe the -1 diffracted SPP mode that strongly interacts with the dipolar LSPR mode for large angles of incidence, while in S-polarization we only observe the dipolar LSPR mode.}
\label{figure3}
\end{figure}

For P-polarized waves the spectral position of the LSPR and SPP modes are switched with respect to the 100 nm disks, which modifies their interaction substantially. The $\nu = -1$ SPP mode is observed at shorter wavelengths than the LSPR mode and it shows the expected red shift with increasing angles. For large angles of incidence, the SPP mode shifts closer to the LSPR mode, which start to interact strongly. The two modes show pronounced anti-crossing behavior, which causes the SPP mode to push the LSPR mode to higher wavelengths. The interaction results in highly asymmetric Fano line shapes for both plasmon modes, where we see the spectral line width increase for the  SPP mode and decrease for the LSPR mode. The LSPR for S-polarized excitation is red shifted compared to the 100 nm disk case and also shows a very minor red shift with increasing angle of incidence. In the next step towards refractive index sensing, we measured the angle dependent reflection spectra of these 140 nm disk samples in water and performed phase sensitive spectroscopic ellipsometry measurements at an incident angle of $30^o$, as presented in figure 4. 

\begin{figure}
\includegraphics[width=8.7cm]{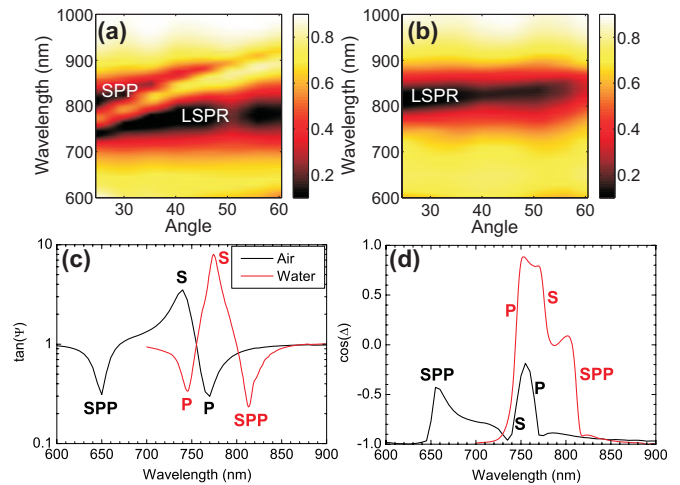}
\caption{Intensity plots for angle dependent reflection measurements on 140 nm disks in water. (a) Measured spectra for P-polarization. (b) Measured spectra for S-polarization. The spectral positions of the resonances for P-polarization have been switched compared to the measurements in air and now clear anti-crossing behavior is observed at small angles of incidence. (c) and (d) Spectroscopic ellipsometry measurements in air and water showing the amplitude reflection ratio $\tan(\Psi)$ and the phase difference $\cos(\Delta)$ between P- and S-polarized reflected waves for an angle of incidens of $30^o$. Clearly in water there is pronounced interaction between the SPP and LSPR-mode in P-polarization, resulting in a large increase of the observed phase differences.}
\label{figure4}
\end{figure}

The amplitude based reflection spectra in water are shown for P- and S-polarized waves in panels a and b respectively. Compared to the measurements in air (figure 3) in P-polarization, we see that the spectral positions of the LSPR and SPP mode have switched. This can be attributed to fact that the decay length for SPPs is much longer than for LSPR modes \cite{OptLettChu}, resulting in a larger bulk sensitivity to the refractive index of the surroundings. Therefore, the SPP mode shifts beyond the LSPR mode, and we see that in this case both modes interact strongly for small angles of incidence, where the SPP mode pushes the LSPR mode to shorter wavelengths (even to shorter wavelengths than for the measurements in air).  Panels c and d show the phase-sensitive spectroscopic ellipsometry data for the same sample measured in air and water at an angle of incidence of $30^o$. The amplitude reflection ratio $\tan \Psi$ shows dips for the localized (P) and propagating (SPP) plasmon modes in P-polarization and a peak for the LSPR in S-polarization (S), while the phase difference $\Delta$ between P- and S-polarized waves is reflected in the $cos \Delta$ value. Similar to our earlier work on randomly distributed nanoparticles \cite{NanoLettLodewijks} we observe a phase difference at the center frequency of the different plasmon modes with a much smaller spectral footprint than for the intensity based measurements. As the effects of inhomogeneous broadening are largely suppressed in our periodic arrays, the spectral footprint of the plasmon resonances is significantly decreased down to 5, 7 and 9 nm for the SPP, P-LSPR and S-LSPR respectively. We observe a very pronounced distinction between the measurements in air and water where the SPP and LSPR modes for the P-polarization show weak and strong interaction respectively. For the measurements in water, where the SPP and LSPR modes show strong Fano-interference and anti-crossing behavior, a pronounced increase in the phase difference is observed for all 3 plasmon modes involved.  Additional spectroscopic ellipsometry data for 100 and 130 nm disks in air and water respectively are provided in the supporting information (S4 and S5), where a pronounced increase of the phase differences is observed as the SPP and LSPR modes show pronounced Fano interference. It is exactly in this region of strong interaction that we performed refractive index sensing measurements, taking benifit both of the strong reduction in line widths and the Fano interference between the LSPR and SPP modes. 

An overview of the ellipsometry based refractive index sensing data is presented in figure 5 and table 1. The samples were mounted in a flow cell and different concentrations of glycerol in water were pumped through in order to investigate the RI sensitivity of the different plasmonic modes. 

\begin{figure}[h]
\includegraphics[width=8.7cm]{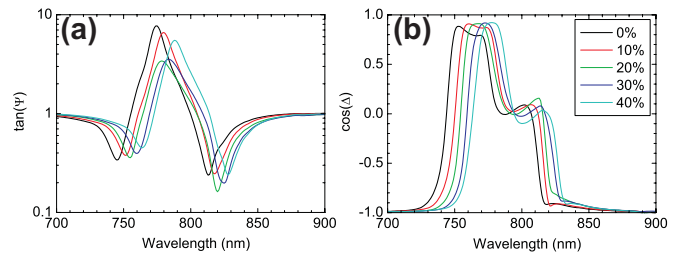}
\caption{Refractive index sensing measurements for different concentrations of glycerol in water with an incident angle of $30^o$. Measured $\tan \Psi$ (a) and $\cos \Delta$ (b) data for 140nm nanodisk samples.}
\label{figure5}
\end{figure}

As expected, we observe a pronounced but different red shift for all the plasmon resonances with increasing values of the refractive index. 

\begin{center}
	\begin{table}
	  \caption{Comparison of the sensitivities and FOMs for refractive index sensing measurement at an incidence angle of $30^{o}$.}
    \begin{tabular}{ | c | c | c | c |}
     \hline
     Mode & P & S & SPP \\ \hline 
     $d\lambda / dn$ (nm/RIU) & 375 & 218 & 291 \\ \hline
     FWHM (nm) & 7 & 9 & 5 \\ \hline 
     FOM & 54 & 24 & 58 \\ \hline
    \end{tabular}
    \end{table}
\end{center}

For the non-interacting S-LSPR we observe a sensitivity comparable to the one obtained for randomly distributed gold nanoparticles \cite{NanoLettLodewijks}, as we would expect. In P-polarization however, we observe quite some interesting and unexpected results. For both plasmon modes, the sensititivity is much higher than for the S-polarized case, which can be attributed to the Fano interference between the SPP and the LSPR mode. As illustrated in figure 4, the SPP mode is observed at longer wavelenths than the LSPR mode, and the interaction between the modes pushes the LSPR to shorter wavelenths than its resonance position in air. When the refractive index of the surroundings is increased, the SPP mode shifts to longer wavelengths and thus it clears spectral space for the LSPR mode, allowing it to also shift closer to its natural resonance position. Therefore we observe a much larger sensitivity for the P-LSPR in the periodic array than for random nanoparticle distributions (almost 2 times larger). This type of Fano-interference can therefore also be exploited to boost the refractive index sensitivity of LSPR modes. If we compare the performance of the periodic samples with randomly distributed nanoparticles, we observed 3 different factors that contribute to a large enhancement of the FOM for RI sensing. First of all, the effects of inhomogeneous broadening are largely reduced, resulting in severe line width reduction in the intensity based measurements. Secondly, by measuring the phase next to the amplitude of the resonances in spectroscopic ellipsometry, an additional reduction of the line width is obtained. Thirdly, the combination with the enhanced RI sensitivity due to the interference of SPP and LSPR modes in P-polarization results in largely enhanced values of the FOM, which reaches values as high as 54, 24 and 58 for the P-LSPR, S-LSPR and SPP mode respectively. 

Surprisingly, the sensitivity for the P-LSPR mode turns out to be higher than the one for the SPP mode and even twice as large as the sensitivity for the S-LSPR. If we compare the sensitivity with the values obtained for randomly distributed nanoparticles \cite{NanoLettLodewijks} , this turns out to be a very unusual result. We would expect the sensitivity of the two LSPR modes to be similar to the ones for the random particles, but we only observe this for the S-LSPR. Moreover we would expect that the sensitivity of the SPP mode would be the largest, as the decay length for a propagating mode is much longer than for a localized mode. In fact, the sensitivity for the SPP mode would be much larger than the one for the P-LSPR if the two modes wouldn't show any pronounced coupling. If we compare the spectra in air and water, we notice that the SPP-mode shifted beyond the P-LSPR mode in water, which already indicates that the SPP mode shows a higher sensitivity to the refractive index. On top of that, for the measurements in water (figure 4 a) we observe anti-crossing behavior between the P-LSPR and SPP mode, which causes the P-LSPR to be blue shifted with respect to its spectral position in air (figure 3 a). When the refractive index of the surroundings is increased, the SPP mode shifts to longer wavelengths and thus away from the P-LSPR mode, allowing this one to also shift closer to its natural resonance position. Therefore we observe a much larger sensitivity for the P-LSPR in the periodic array than for random nanoparticle distributions. This type of Fano-interference can therefore also be exploited to boost the refractive index sensitivity of LSPR modes. If we compare the performance of the periodic samples with randomly distributed nanoparticles, we observed 3 different factors that contribute to a large enhancement of the FOM for RI sensing. First of all, the effects of inhomogeneous broadening are largely reduced, resulting in severe line width reduction in the intensity based measurements. Secondly, by measuring the phase next to the amplitude of the resonances in spectroscopic ellipsometry, an additional reduction of the line width is obtained. Thirdly, the combination with the enhanced RI sensitivity due to the interference of SPP and LSPR modes in P-polarization results in largely enhanced values of the FOM, which reaches values as high as 54, 24 and 58 for the P-LSPR, S-LSPR and SPP mode respectively. 

To summarize, we have shown that the interaction between localized and propagating plasmon resonances in periodic arrays of gold nanoparticles on top of a silica spacer and a continuous gold layer can be used to tune the refractive index sensing performance of the LSPR. By adjusting the size of the nanoparticles and the pitch it is possible to tailor the optical response in such a way that the localized and propagating modes can be excited in a small spectral window. The Fano-interference between these 2 modes results in more pronounced phase differences with reduced line widths, making them very useful for refractive index sensing applications. The sensitivity of the LSPR can be enhanced when it interacts with the SPP mode and can reach values which are twice as large as those for the non-interacting mode. The resulting line widths range between 5 and 10 nm and the FOM values reach values in between 24 and 58 for the different plasmon modes.

{\bf Supporting information: } Additional information on sample fabrication, coupling to SPP grating modes, comparison between randomly distributed and periodic arrays of nanodisks and additional optical spectra for different nanodisk sizes. This material is available free of charge via the Internet at http://pubs.acs.org.

{\bf Acknowledgements: } The authors thank J. Moonenes, E. Vandenplas, K. Baumans and J. Feyaerts for technical support. K.L. Acknowledges IWT Flanders and P.V.D. acknowledges FWO Flanders for financial support.


\begin{thebibliography}{24}

\bibitem{NatMaterAnker}
Anker J. N., Paige Hall W., Lyandres O., Shah N. C., Zhao J. and Van Duyne R. P., 
Biosensing with plasmonic nanosensors.
\newblock {\it Nat. Mater.} {\bf 7}, pp. 442-453 (2008)

\bibitem{NLSvedendahl2009}
Svedendahl M., Chen S., Dmitriev A. and K\"all M.,
Refractometric Sensing Using Propagating versus Localized Surface Plasmons: A Direct Comparison.
\newblock {\it Nano Lett.} {\bf 9}, pp. 4428-4433 (2009)

\bibitem{OptExprPiliarik}
Piliarik M., Kvasnička P., Galler N., Krenn J. R., and Homola J.,
Local refractive index sensitivity of plasmonic nanoparticles.
\newblock {\it Opt. Expr.} {\bf 19}, pp 9213-9220 (2011)


\bibitem{APLYe}
Ye J., Shioi M., Lodewijks K., Lagae L., Kawamura T. and Van Dorpe P.,
Tuning plasmonic interaction between gold nanorings and a gold film for surface enhanced Raman scattering.
\newblock {\it Appl. Phys. Lett.} {\bf 97}, pp 163106 (2010)

\bibitem{NanoLettYe}
Ye J., Wen F., Sobhani H., Lassiter J. B., Van Dorpe P., Nordlander P. and Halas N. J., 
Plasmonic Nanoclusters: Near Field Properties of the Fano Resonance Interrogated with SERS
\newblock {\it Nano Lett.} {\bf 12} (3), pp. 1660-1667 (2012)


\bibitem{ACSOffermans}
Offermans P.,Schaafsma M. C., Rodriguez S. R. K., Zhang Y., Crego-Calama M., Brongersma S. H., and Gomez Rivas J., 
Universal Scaling of the Figure of Merit of Plasmonic Sensors.
\newblock {\it ACS Nano} {\bf 5}, pp. 5151-5157 (2011)

\bibitem{NanoTechChen}
Chen S., Svedendahl M., K\"all M., Gunnarsson L. and Dmitriev A., 
Ultrahigh sensitivity made simple: nanoplasmonic label-free biosensing with an extremely low limit-of-detection for bacterial and cancer diagnostics.
\newblock {\it Nanotechn.} {\bf 20}, pp. 434015 (2009)


\bibitem{ACSFrancescato}
Francescato Y., Giannini V. and Maier S. A., 
Plasmonic Systems Unveiled by Fano Resonances
\newblock{\it ACS Nano} {\bf 6} (2), pp. 1830-1838 (2012)

\bibitem{ACSSvedendahl}
Svedendahl M. and K\"all M., 
Fano Interference between Localized Plasmons and Interface Reflections.
\newblock{\it ACS Nano} {\bf 6} (8) pp. 7533 (2012)

\bibitem{NanoLettVerellenDolmen}
Verellen N., Sonnefraud Y., Sobhani H., Hao F., Moshchalkov V.V., Van Dorpe P., Nordlander P. and Maier S. A. 
Fano Resonances in Individual Coherent Plasmonic Nanocavities
\newblock {\it Nano Lett.} {\bf 9} (4), pp. 1663-1667 (2009)


\bibitem{ACSHao}
Hao F., Nordlander P., Sonnefraud Y., Van Dorpe P., Maier S. A., Tunability of Subradiant Dipolar and Fano-Type Plasmon Resonances in Metallic Ring/Disk Cavities: Implications for Nanoscale Optical Sensing.
\newblock {\it ACS Nano} {\bf 3}, pp. 643-652 (2009)

\bibitem{NanoLettVerellenCross}
Verellen N., Van Dorpe P., Huang C., Lodewijks K., Vandenbosch G. A. E., Lagae L. and Moshchalkov V.V., Plasmon Line Shaping using Nanocrosses for High sensitivity LSPR Sensing.
\newblock {\it Nano Lett.} {\bf 11} (2), pp. 391-397 (2011)

\bibitem{NanoLettHao}
Hao F., Sonnefraud Y., Van Dorpe P., Maier S. A., Halas N. J., Nordlander P., 
Symmetry Breaking in Plasmonic Nanocavities: Subradiant LSPR Sensing and a Tunable Fano Resonance.
\newblock {\it Nano Lett.} {\bf 8}, pp. 3983-3988 (2008)

\bibitem{NatMaterLulyanchuk}
Luk'yanchuk B., Zheludev N. I., Maier S. A., Halas, N. J., Norlander P, Giessen H and Chong T. C., 
The Fano resonance in plasmonic nanostructures and metamaterials. 
\newblock {\it Nat. Mater.} {\bf 9}, pp. 707-715 (2010)

\bibitem{NanoLettEvlyukhin}
Evlyukhin A. B., Bozhevolnyi S. I., Pors A., Nielsen M. G., Radko I. P., Willatzen M. and Albrektsen O., Detuned Electrical Dipoles for Plasmonic sensing.
\newblock {\it Nano Lett.} {\bf 10}, pp. 4571-4577 (2010)


\bibitem{OptCommKabashin}
Kabashin A.V. and Nikitin P.I., 
Surface plasmon resonance interferometer for bio- and chemical-sensors.
\newblock {\it Opt. Comm.} {\bf 150}, pp. 5-8 (1998)

\bibitem{OptExprKabashin}
Kabashin A.V., Patskovsky S. and Grigorenko A. N., 
Phase and amplitude sensitivities in surface plasmon resonance bio and chemical sensing.
\newblock {\it Opt. Expr.} {\bf 17}, pp. 21191-21204 (2009)

\bibitem{NanoLettLodewijks}
Lodewijks K., Van Roy W., Borghs G., Lagae L. and Van Dorpe P., 
Boosting the Figure-Of-Merit of LSPR-based Refractive Index Sensing by Phase-Sensitive Measuremnts.
\newblock {\it Nano Lett.} {\bf 12} (3), pp. 1655-1659 (2012)

\bibitem{OptLettKravets}
Kravets, V. G., Schedin, F., Kabashin, A. V. and Grigorenko, A. N., Sensitivity of Collective Plasmon Modes of Gold Nanoresonators to Local Environment.
\newblock {\it Opt. Lett.} {\bf 35}, pp. 956-958 (2010)


\bibitem{OptLettChu}
Chu Y. and Crozier K. B.,
Experimental study of the interaction between localized and propagating surface plasmons.
\newblock {\it Opt. Lett.} {\bf 34}, pp. 244-246 (2009)

\bibitem{APLGhoshal}
Goshal A., Divliansky I. and Kik P. G.,
Experimental observation of mode-selective anti-crossing in surface-plasmon-coupled metal nanoparticle arrays.
\newblock {\it Appl. Phys. Lett.} {\bf 94}, pp. 171108 (2009)

\bibitem{PRBChrist}
Christ A., Zentgraf T., Tikhodeev S. G., Gippius N. A., Kuhl J. and Giessen H.,
Controlling the interaction between localized and delocalized surface plasmon modes: Experiment and numerical calculations.
\newblock {\it Phys. Rev. B.} {\bf 74}, pp 155435 (2006)


\bibitem{SOPRA}
GESP 5 ellipsometer, SOPRALAB, 55, avenue de l'Europe, 92400 COURBEVOIE, France

\bibitem{NanoLettChen}
Chen, K-P., Drachev V. P. Borneman J. D. Kildishev A. V. and Shalaev V. M.
Drude relaxation rate in grained gold nanoantennas.
\newblock {\it Nano Lett.} {\bf 10} (3), pp. 916-922 (2010)

\bibitem{COMSOL}
Comsol Multiphysics 3.5a (RF Module), 1 New England Executive Park, Suite 350, Burlington, MA 01803, USA




\end{thebibliography}
\end{document}